\documentclass{aa}
\usepackage{psfig}
\begin{document}

   \thesaurus{
               09.08.1;  
               09.09.1 G9.62+0.19;  
               08.06.2}  
\newcommand{\G}{G9.62+0.19}  
\newcommand{\hii}{H{\sc ii}}    
\title{Detection of the thermal radio continuum emission from  the \G-F Hot
       Core}


\author{L. Testi \inst{1}
        \and P. Hofner\inst{2,3}
        \and S. Kurtz\inst{4}
        \and M. Rupen\inst{5}
       }

\offprints{L. Testi; lt@arcetri.astro.it}

\institute{Osservatorio Astrofisico di Arcetri,
           Largo E. Fermi 5, I-50125 Firenze, Italy
           \and
           Physics Department, University of Puerto Rico at Rio Piedras, 
           P.O. Box 23343, San Juan, Puerto Rico 00931
           \and
           Arecibo Observatory, NAIC/Cornell University, HC3 Box 53995, 
           Arecibo, Puerto Rico 00612
           \and
           Instituto de Astronom\'{\i}a, UNAM, Apdo. Postal 70-264, 
           04510 M\'exico D.F., M\'exico
           \and
           National Radio Astronomy Observatory, Socorro, NM 87801
           }

\date{}

\maketitle

\begin{abstract}

We present new high resolution and high sensitivity multi-frequency VLA
radio continuum
observations of the \G-F hot molecular core. We detect for the first time 
faint centimetric radio continuum emission at the position of the core.
The centimetric continuum spectrum of the source is consistent with
thermal emission from ionised gas.
This is the first direct evidence that a newly born massive star is
powering the \G-F hot core.

   \end{abstract}

%

\section{Introduction}
\label{sintro}

The formation of massive stars (M$\ge 10$~M$_\odot$) has received growing 
attention in recent years, because of their important role in galactic 
evolution and the recognition that 
the majority of low-mass stars
are formed together with high-mass stars in clusters (Clarke et 
al.~\cite{C00}).
One of the earliest manifestations of a newly 
born massive star is the appearance of an ultracompact (UC) HII region 
produced by the strong UV stellar radiation field. Since most massive stars 
are formed in clusters it is expected that other forming massive stars can 
be found close to UCHIIs. Indeed, NH$_3$(4,4) high angular resolution 
observations of the molecular environment around UCHIIs (Cesaroni et 
al.~\cite{Cea94}) revealed compact (size $\sim 0.1$~pc) and high 
temperature (T$_{kin}\sim 100$~K) molecular clumps, so-called 
hot cores (HCs). HCs are indeed close, but not generally coincident with, 
the UCHII (Cesaroni et al.~\cite{Cea94}). Given the high energy input 
required to maintain the HCs at the observed temperature, they are likely
to be heated by young high mass stars. Since massive stars are expected
to reach the main sequence while still accreting (Palla \& Stahler~\cite{PS93}),
the lack of centimeter radio continuum emission at a few mJy level 
(e.g. Cesaroni et al.~\cite{Cea94}) can be explained in terms of
the confinement provided by the pressure of the hot molecular gas or
by the 
infalling material accreting onto the
massive star (e.g. de~Pree et al.~\cite{dPea95};
Xie et al.~\cite{Xea96}; Walmsley~\cite{Wmex}).
Either of these could effectively block the expansion of the
ionised gas and make the radio continuum emission extremely compact, optically
thick, and thus not easily detectable at cm wavelengths.
HCs have been suggested to be sites of massive star formation
(Cesaroni et al.~\cite{Cea94}; Kurtz et al.~\cite{K00}). 
However, if a young massive star is indeed present 
inside the HC and is heating the molecular gas, a region of ionised
gas should be present around the star, albeit compact. The
cm radio emission from these objects should be optically 
thick and unresolved, with emission measures exceeding 10$^8$~cm$^{-6}$\,pc.

At a distance of 5.7~kpc (Hofner et al.~\cite{Hea94}), \G\
is a well known UCHII complex extensively studied at high resolution in
the centimetric radio continuum (e.g. Garay et al.~\cite{Gea93}; Cesaroni et
al.~\cite{Cea94}, among others). Several HII and UCHII regions in different 
evolutionary phases are present in the region, and there are indications
of a possible age gradient going from the western, older, regions toward the
eastern, younger, ones (Hofner et al.~\cite{Hea94}; \cite{Hea96};
Testi et al.~\cite{Tea98}).
The centimetric radio continuum components have been designated from A to E
(Garay et al.~\cite{Gea93}). High resolution thermal molecular line
and millimeter continuum observations revealed the presence of a HC in
close coincidence with maser emission from several different
molecules and located midway between radio components D and E.
This new component, without detected centimeter continuum emission but
associated to hot, $\sim$100~K, NH$_3$, CH$_3$CN and dust emission,
was called F (Cesaroni et al.~\cite{Cea94}; Hofner et al.~\cite{Hea94}, 
\cite{Hea96}).
Inside the molecular hot core a young massive star is presumed to be 
forming (Hofner et al.~\cite{Hea96}; Testi et al.~\cite{Tea98}).
If a young massive object is indeed present within the 
HC, an order-of-magnitude calculation suggests that the free-free
radio continuum emission at 22~GHz should be optically thick with a 
total flux of $\sim 0.2$--$0.6$~mJy and with a spatial extent of $\sim 10$
mas (Testi et al.~\cite{Tea98}).
We thus decided to perform VLA high sensitivity radio continuum observations
to detect the faint centimetric free-free emission
expected from such an object.

\section{Observations}
\label{sobs}

The \G\ region was observed with the NRAO\footnote{The 
National Radio Astronomy Observatory is operated by Associated 
Universities, Inc., under contract with the National Science Foundation.} 
VLA in the period May-June~1998 in the radio continuum 
at 3.6 and 1.3~cm, and on 26 January 1999
at 0.7 and 2~cm. The observing parameters are summarized in 
Table~\ref{tobs}. The 2~cm dataset was obtained as a byproduct of the 
0.7~cm experiment: all antennas equipped with Q-band 
receivers available at the time of the observations (12) were
used at 0.7~cm while the remaining 15 were employed at 2~cm.
At 0.7~cm we used a fast-switching observing cycle with 
80~s on-source and 40~s on-calibrator ($\sim$6$^\circ$ away),
which resulted in a total switching cycle of $\sim$160~s and an
efficiency of $\sim$50\%. Hourly pointing sessions 
on the phase calibrator at 3.6~cm were used to correct for 
pointing drifts at 1.3 and 0.7~cm. 3C286 and/or 3C48 
were observed to set the flux scale, which is expected to be
accurate within 10--15\%.

All data editing, calibration and imaging were performed 
within the AIPS software package. After standard flux and 
complex gain calibration, each dataset was self-calibrated using
one phase-only and one phase and amplitude
iteration. Consistency among maps at
different frequencies provided an internal consistency check
of our calibration procedures. Comparison of our fluxes with
previous observations for component D provided an additional check. 
At 0.7~cm heavy data editing was required due to poor atmospheric 
conditions, only $\sim$12\% of the entire dataset was used to produce 
the final maps, corresponding to the last $\sim$30 minutes of 
the run when atmospheric fluctuations settled.
All maps presented here have been obtained using the AIPS IMAGR task with 
uniform weighting of the visibilities and with the ROBUST parameter set to zero.
In all cases we imaged an area at 
least equal to the primary beam FWHM (see Table~\ref{tobs}) to search for 
emission. No correction for primary beam attenuation has been applied at 
any frequency.

\begin{table}
\caption[]{\label{tobs}VLA observing parameters}
\begin{tabular}{rcc}
\hline
Phase center & \multicolumn{2}{c}{$\alpha$(J2000)=18$^h$06$^m$14.88$^s$}\\
             & \multicolumn{2}{c}{$\delta$(J2000)=$-$20$^\circ$31$^\prime$40.8$^{\prime\prime}$}\\
\hline
Parameter & 3.6~cm & 1.3~cm \\ 
\hline
Date     & 12May/05Jun98 & 15/17Jun98 \\ 
Configuration & A/BnA& BnA \\ 
Time on source (hrs)& 0.5/1.0 & 2.0/4.0 \\ 
Freq./Bandwidth (GHz) &8.46/4$\times$0.05 &22.46/4$\times$0.05 \\ 
Flux cal 3C286/3C48 (Jy) & 5.18/-- & 2.5/1.17 \\ 
Phase cal 1820$-$254 (Jy) & 1.0 & 1.0 \\ 
Primary beam (FWHP) &5.$\!^\prime$4 &2\arcmin \\ 
Largest structure  &10\arcsec &5\arcsec \\ 
Syn. beam (FWHM;pa) & $0\farcs 37\times 0\farcs 24$;$-$5$^\circ$ & 
            $0\farcs 28\times 0\farcs 18$;64$^\circ$
          \\ 
Noise (mJy/beam)        &0.018 &0.062 \\ 
\hline
Parameter & 2~cm & 0.7~cm \\
\hline
Date     & 26Jan99 & 26Jan99 \\
Configuration &  C & C \\
Time on source (hrs)& 4.5 & 2.5 \\
Freq./Bandwidth (GHz) &14.94/4$\times$0.05 &43.34/4$\times$0.05 \\
Flux cal 3C286/3C48 (Jy) &  3.43/-- & 1.45/0.57 \\
Phase cal 1820$-$254 (Jy) & 0.87 & 0.86\\
Primary beam (FWHP) &3\arcmin &1\arcmin \\
Largest structure  & 30\arcsec & 10\arcsec \\
Syn. beam (FWHM;pa) &$2\farcs 4\times 1\farcs 5$;5$^\circ$ &$0\farcs 68\times 0\farcs 64$;$-$33$^\circ$ \\
Noise (mJy/beam)        &0.1 &1.0 \\
\hline
\end{tabular}
\vskip -0.35cm
\end{table}

\section{Results}
\label{sres}

\begin{figure*}
\centerline{\psfig{figure=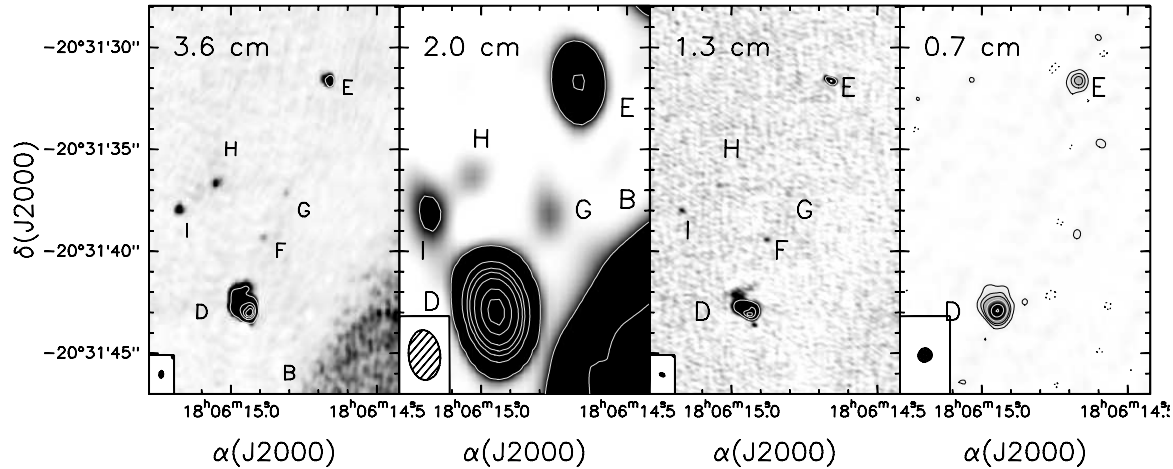,width=15.5cm,angle=-90}}
\caption[]{\label{fcont}3.6, 2.0, 1.3, and 0.7~cm radio continuum images of the
\G D--I complex. The detected sources are labelled from D to I (sources A,
 and C are outside the region shown, source B is only partially visible). 
At each wavelength, the synthesised beam FWHM (see
Table~\ref{tobs}) is shown by the ellipses in the lower left corner of each
panel. 
}
\vskip -0.35cm
\end{figure*}

In Figure~\ref{fcont} we show our radio continuum images at 3.6, 2.0, 1.3,
and 0.7~cm of the region containing the known cm-continuum components D--E
(e.g. Cesaroni et al.~\cite{Cea94}). Components A, B, and C (Garay et
al.~\cite{Gea93}) are detected in some of our maps depending on sensitivity
and ($u,v$) coverage, and are outside the shown area.

In addition to all the previously known cm-continuum components, at 3.6
and 1.3~cm we 
detect four additional sources, labelled F to I, all above the 8$\sigma$
level of $\sim 0.14$~mJy/beam at 3.6~cm. At 2.0~cm only component F is not
detected.
In Table~\ref{tspar}, 3.6~cm peak positions and integrated fluxes or upper 
limits at each frequency for each of the newly detected radio continuum
components are reported; all the newly detected sources are unresolved by our
synthesised beams. A detailed study of all the detected sources goes
beyond the scope of the present letter and will be presented in a
forthcoming paper.

\begin{figure}
\centerline{\psfig{figure=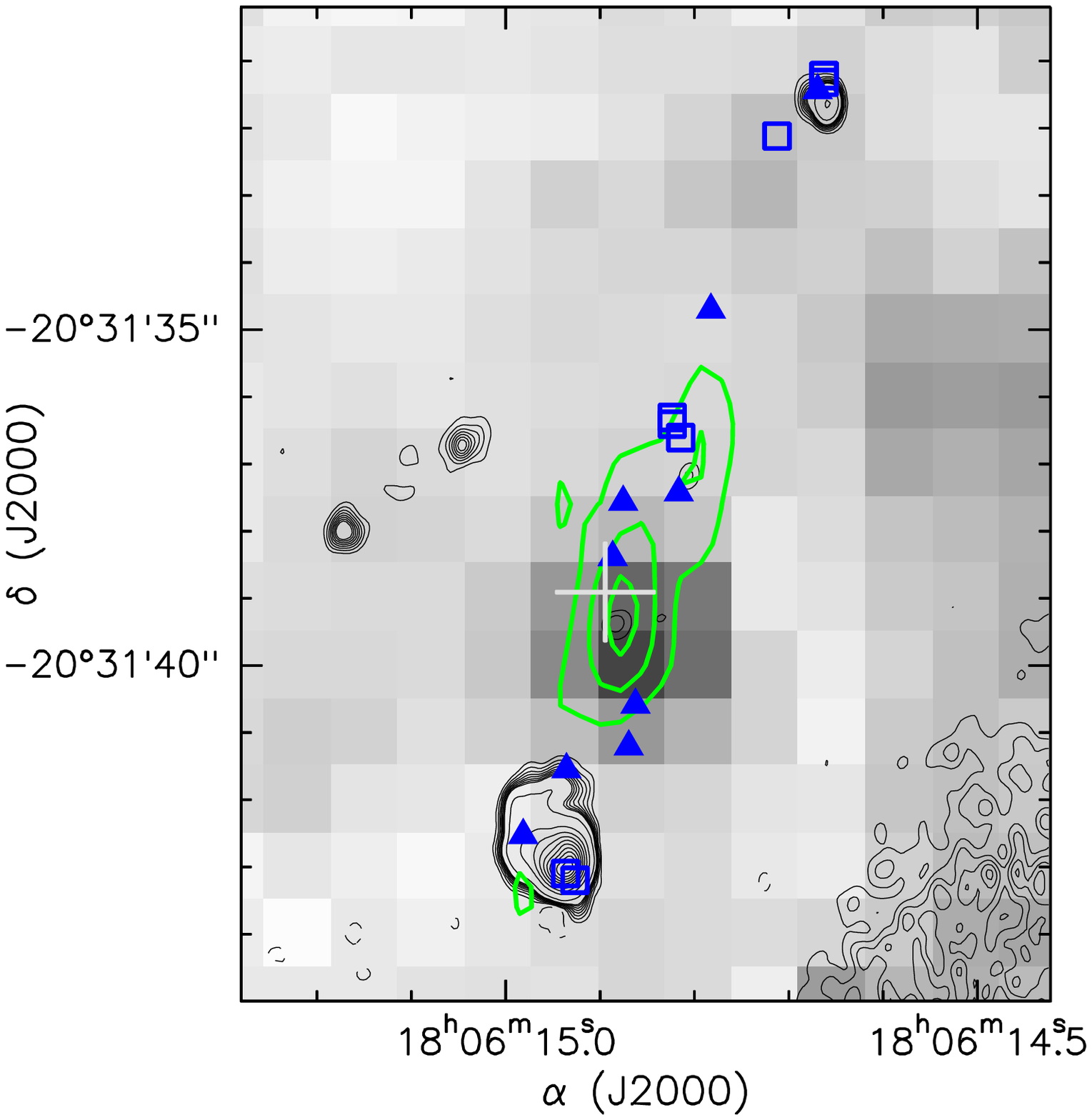,width=6.7cm,angle=-90}}
\caption[]{\label{fcontq}Our 3.6~cm continuum (thin contours) and
the thermal NH$_3$(5,5) emission (thick contours, Hofner et al.~\cite{Hea94})
maps are overlaid on the 2.2\,$\mu$m near infrared image of Testi et
al.~(\cite{Tea98}) of the \G D--F region.
Filled triangles are H$_2$O masers (Hofner \& Churchwell~\cite{HC96}),
open squares OH masers (Forster \& Caswell~\cite{FC89}), the
cross marks the position of the mm continuum source F
(Hofner et al.~\cite{Hea96}; CH$_3$CN and mm-continuum emission is also found
in coincidence with sources D and E, but are not shown in this figure to
avoid confusion).
The 3.6~cm contour levels are 
--50, 50 to 200 by 50~$\mu$Jy/beam, 0.3 to 0.9 by 0.1~mJy/beam,
and 2 to 18 by 4~mJy/beam. The NH$_3$(5,5) contours are --0.14, 0.14 to 0.3 by
0.77~Jy\,km/s/beam.
} 
\vskip -0.35cm
\end{figure}

In Figure~\ref{fcontq} we show the position of the H$_2$O and OH masers
(Forster \& Caswell~\cite{FC89}; Hofner \& Churchwell~\cite{HC96}),
and of the mm-continuum component F (Hofner et al.~\cite{Hea96})
and our 3.6~cm (thin contours) and the thermal NH$_3$(5,5) (thick 
contours, Hofner et al.~\cite{Hea94}) maps
overlaid on the 2.2~$\mu$m near infrared image from Testi et al.~(\cite{Tea98}).
Within the astrometric uncertainties ($\le$1\arcsec\ for the NIR data,
$\le$0.2\arcsec\ for all the other data), the newly discovered cm-continuum
source called F in Table~\ref{tspar} is coincident with the NH$_3$(5,5) HC, the
mm-continuum and the NIR source. 

\begin{table}
\caption[]{\label{tspar}Observed parameters of the newly detected sources}
\begin{tabular}{lccrrrr}
\hline
& $\alpha$(J2000) & $\delta$(J2000) & F$_{3.6\rm cm}$ & F$_{2.0\rm cm}$ & F$_{1.3\rm cm}$ & F$_{0.7\rm cm}$\\
      & 18$^h$06$^m$& $-$20$^\circ$31\arcmin & (mJy) & (mJy) & (mJy) & (mJy) \\
\hline
F & 14.$\!^s$884 & 39\farcs37 & 0.22   & $<$0.5 & 0.62    & $<$3 \\
G & 14.$\!^s$805 & 37\farcs17 & 0.15   & 0.7$^a$& 0.39    & $<$3 \\
H & 15.$\!^s$047 & 36\farcs72 & 0.98   & 0.4    & 0.36    & $<$3 \\
I & 15.$\!^s$172 & 37\farcs97 & 1.0    & 1.6    & 0.8     & $<$3 \\
\hline
\end{tabular}
\vskip 0.1cm
{\small 
$^a$) The 2~cm position of source G is not exactly coincident with that
measured at 3.6 and 1.3~cm; the faint source is very close to bright and
extended sources, the flux cited is thus highly uncertain due to possible
imaging artifacts. Only higher angular resolution observations at 2~cm will 
offer a more accurate position and flux.}
\vskip -0.35cm
\end{table}


\begin{figure}
\centerline{\psfig{figure=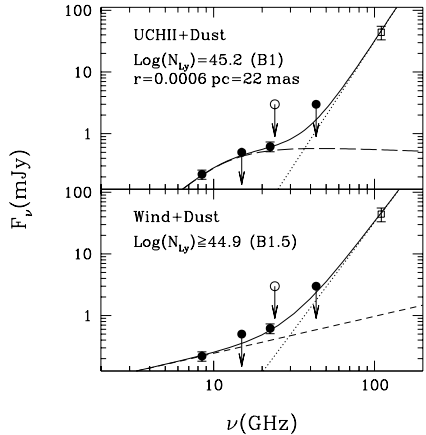,width=6.0cm}}
\caption[]{\label{fspec}Radio spectrum of component F. Filled circles:
3.6, 2.0, 1.3 and 0.7~cm this work;
open circle: 1.3~cm from 
Cesaroni et al.~(\cite{Cea94}); open square: 2.7~mm from Hofner et
al.~(\cite{Hea96}). Upper limits are indicated by arrows.
Top panel: spherical homogeneous UCHII region (long dashed line) plus
thin dust emission (dotted line; $\beta$=2). Bottom panel: spherical
wind (short dashed line) plus thin dust emission (dotted line; $\beta$=1.5).}
\vskip -0.35cm
\end{figure}

The HC and mm-component F located between the cm-continuum sources
D and E was the primary target of our observations.
We detected source F at 3.6 and 1.3~cm and set 
upper limits at 2.0 and 0.7~cm.
In Figure~\ref{fspec} we show
the radio continuum spectrum of source F.
Data from the present work
are presented as filled circles, while
the open circle is from Cesaroni et al.~(\cite{Cea94}), and the
open square from Hofner et al.~(\cite{Hea96}).

\section{Discussion and Conclusions}
\label{sdis}

The primary goal of our new VLA observations was the detection
of radio continuum emission from the mm-component and NIR source F, as
predicted by simple considerations in the case that a massive
young stellar object is hidden within and is heating the HC
(Testi et al.~\cite{Tea98}). As shown in the previous section,
one of our newly detected cm-continuum sources is coincident with the 
HC. In Fig.~\ref{fspec} we show the radio continuum spectrum of the HC.
The HC emission can be well fitted by a two
component model: free-free emission from ionised gas plus optically thin 
thermal dust emission at mm-wavelengths.
The presence of ionised gas requires a continuum source of energy either in
the form of a UV photoionization field or collisional ionisation.
The detection of warm dust is consistent with the presence of a very young
massive star, as implied also by the molecular gas observations (Cesaroni
et al.~\cite{Cea94}; Hofner et al.~\cite{Hea96}).

The ionised gas could be distributed either within a very compact
($r\sim0.0006$\,pc)
spherical homogeneous UCHII with EM$\sim$3$\times$10$^8\rm\,cm^{-6}\,pc$,
or in a spherical wind (or $r^{-2}$ density gradient)
with F$_\nu\propto\nu^{0.6}$ 
(Panagia \& Felli~\cite{PF75}), in either
case the Lyman photon supply rate could be provided by a 
zero age main sequence star earlier than B1-B1.5 (Panagia~\cite{P73}).
This is in agreement with independent estimates of the expected spectral type
derived from the molecular and infrared observations (O9-B0.5;
Hofner et al.~\cite{Hea96}; Testi et al.~\cite{Tea98}).
The optically thin thermal dust emission requires a dust
emissivity index in the range $\beta$=1.5-2.0.

Only two other HCs have been detected in the cm continuum
W3(H$_2$O) (Reid et al.~\cite{Rea95}; Wilner et al.~\cite{Wea99})
and IRAS~20126$+$4104 (Hofner et al.~\cite{Hea99}).
In the first case the radio emission is non-thermal from a synchrotron
jet, thus no constraint can be obtained on the nature of the central
(proto-)star. IRAS~20126$+$4104
has been detected at one frequency only in the cm continuum,
and the mm spectrum is consistent with pure dust emission, while
the cm emission is most probably due to a jet, the exact nature of
which has not been firmly established yet.
In the case of \G-F the thermal origin of the emission is confirmed by the 
radio spectral index constraint. The source is unresolved in our $\sim$200~mas
($\sim 1200$~AU)
beam, while the two jets observed in the other cores, scaled to the distance
of \G, would have been marginally resolved by our observations. Nevertheless,
the possibility of a thermal jet cannot be completely ruled out.

An alternative interpretation for HCs is that
they could be molecular clumps that are heated from the outside
by a nearby star. This scenario is unlikely on theoretical grounds
(Kaufman, Hollenbach and Tielens~\cite{KHT98}) and 
the radial temperature profiles which were measured for three different
HCs by Cesaroni et al.~(\cite{Cea98}) indicate internal heating in those cases.
In the case of G9.62+0.19--F, our data provide plausible (but not decisive)
arguments for internal heating. First, the continuum source and the
peak of the molecular gas as given by NH$_3$(5,5) coincide to very
high accuracy (better than $0\farcs3$). This would not necessarily be
the case for external heating by an unrelated star outside the HC.   
Second, the measured brightness temperature of the continuum of $42\,$K,
together with the indication of appreciable continuum optical depth implies a
very small linear size of the emitting region of about $120\,$AU. 
Thus, even if the object that is ionising source F is external, it must be
very nearby because the fraction of ionising photons emitted into the
solid angle subtended by component F scales as $\left (D\over r\right )^2$,
where $r$ is the distance to the illuminating source and $D$ the clump size
of 120~AU.  For instance, at a 
distance of $1200\,$AU the required ZAMS spectral type of
the illuminating star would be O9.5 and it is very unlikely that the effects
of such a star on the surrounding matter would remain undetected.

In summary, the presence of unresolved, optically thick, thermal
free-free emission is strong direct evidence for a newly born massive star
within the \G-F HC. Thus, even within this complex cluster of
UCHII regions, the heating of the hot molecular gas within the HC
is most probably produced by an embedded young massive star.
This is consistent
with the idea that the HC phase is an evolutionary phase of young massive
stars preceeding the formation of an UCHII. However, in order to make 
a firm statement in this respect, a larger sample of HCs should be observed
at several frequencies in the cm radio continuum at high resolution and
sensitivity.

\begin{acknowledgements}
We thank Riccardo Cesaroni, Marcello Felli and the referee, Todd Hunter,
for very useful comments and stimulating
discussion, and Barry Clark for nice scheduling at the VLA. 
LT was partially supported by
NASA Origins of the Solar System program through grant NAGW-4030.
\end{acknowledgements}

\end{document}